\documentclass[aps,amsmath,amssymb,twocolumn,floatfix,showpacs,showkeys]{revtex4-1}
\usepackage{graphicx}
\usepackage{natbib}
\usepackage{dcolumn}
\usepackage{bm}
\usepackage{hyperref}
\usepackage{xcolor}

\def\defeq{\mathrel{\mathop:}=}

\begin{document}
\newcolumntype{C}{>{\centering\arraybackslash}p{2em}}


\title{Energetic, electronic and magnetic properties of Mn-pairs on reconstructed (001) GaAs surfaces.}


\author{Magdalena Birowska}
\email{Magdalena.Birowska@fuw.edu.pl}
\affiliation{Faculty of Physics, University of Warsaw, Pasteura 5, PL-02-093 Warszawa, Poland,}
\author{Cezary \'{S}liwa}
\affiliation{Institute of Physics, Polish Academy of Sciences, al. Lotnik\'{o}w 32/46, PL-02-668 Warszawa, Poland,}
\author{Jacek A. Majewski}
\affiliation{Faculty of Physics, University of Warsaw, Pasteura 5, PL-02-093 Warszawa, Poland.}


\date{\today}

\begin{abstract}
We study energetic, magnetic, and electronic properties of diluted substitutional Mn-pairs on the reconstructed $(001)$ GaAs surfaces. The studies are based on first-principles calculations in the framework of the density functional theory. We demonstrate that the stability of the systems strongly depends on the position, orientation, and the distance between the Mn-atoms constituting the pair. Independently of the considered surface reconstruction pattern, the Mn-pairs with Mn-atoms being the nearest neighbors (NN) on cationic sublattice  turn out to be energetically more favorable than the pairs with the larger distance between the Mn-atoms. However, the preferential build-up orientation of the Mn-NN-pair depends on the surface reconstruction and is parallel either to $[110]$ or $[1\bar{1}0]$ crystallographic direction. We reveal also the mechanisms of the magnetic ordering of Mn-NN-pairs. The Mn-NN-pairs along the $[110]$ crystallographic direction exhibit always ferromagnetic alignment of Mn spins, whereas the spins in the Mn-NN-pairs along $[1\bar{1}0]$ direction are mostly anti-ferromagnetically aligned. In the electronic structure of the systems containing Mn-pairs with ferromagnetically aligned spins, we observe the valence band hole states in the neighborhood of Fermi energy. This indicates that the surface ferromagnetism in this prototype of dilute magnetic semiconductors can be explained in terms of the $p$-$d$ Zener model.
\end{abstract}



\maketitle

\section{Introduction}
Nowadays, the epitaxially grown semiconducting magnetic films like (Ga,Mn)As have attracted great deal of attention, mostly due to its intriguing physical properties, such as magnetocrystaline anisotropy allowing for the control of magnetotransport phenomena \cite{RevModPhys.86.187,RevModPhys.86.855}, that make them promising candidates for spintronic applications. The properties of these materials depend strongly on how the magnetic ions are incorporated into the host. For example, it is commonly accepted \cite{RevModPhys.86.187,RevModPhys.86.855} that the interstitial Mn-ions in bulk GaAs reduce the concentration of holes in the valence band, and hence diminish the Curie temperature and hinder \textit{p-d } Zener type ferromagnetism. On the other hand, it was predicted \cite{PhysRevB.68.075202} that interstitial Mn-ions might induce new type of ferromagnetism in Mn-doped GaAs through formation of the substitutional-interstitial Mn complexes.  Moreover, it has been demonstrated by the authors \cite{PhysRevLett.108.237203} that the bulk uniaxial, in-plane and out-of-plane, magnetic anisotropies originate from the existence of the preferential buildup direction of the Mn-atoms being the nearest neighbors along a crystallographic direction. This preferential build-up direction of Mn-atoms has been ascribed to the growth mechanism of (Ga,Mn)As and the manner the Mn-pairs incorporate into the (001) GaAs surface during the growth process \cite{PhysRevLett.108.237203}. 

Since the uniaxial magnetic anisotropy in these dilute magnetic semiconductors (DMSs) is of crucial importance for design of novel spintronics devices based on these systems \cite{Nature.455.2008}, the deep physical understanding of its origins and also finding out the mechanisms that allow for tuning the anisotropy through the suitable choice of growth conditions are worth of further studies. Furthermore, it has been recently reported that a single atom substitution technique together with spectroscopic imaging in a scanning tunneling microscope (STM) open a new way to manipulate atom by atom at the surface \cite{Nature.442.2006}, and therefore, the surface magnetism. In the present studies, we employ {\sl ab initio} calculations in the framework of the density functional theory (DFT) to address the following issues: (i) the energetics of the Mn-pair substituted into the cationic sublattice on various reconstructed (001) GaAs surfaces, (ii) the influence of the local environment on relative orientation of localized spins of Mn-atoms and induced type of surface magnetism, (iii) the determination of the exchange constants for Mn-Mn interactions in spin Hamiltonians, (iv) the changes of the surface electronic structure induced by the incorporated Mn-atoms. We aim to study the interactions between two Mn-atoms unaffected by the presence of the other magnetic atoms, so we consider the lowest computationally tractable Mn coverage of the surface (12.5\%), which corresponds roughly to the highest Mn concentration in the synthesized (Ga:Mn)As bulks \cite{RevModPhys.86.187,RevModPhys.86.855}. To determine the physical model for our studies, we have made use of findings reported on the basis of previous DFT calculations for GaAs bulk doped with Mn \cite{PhysRevB.68.075202}, and scanning tunneling microscopy (STM) measurements supported by DFT calculations for Mn covered surfaces \cite{PhysRevLett.89.227201,PhysRevB.69.121308,PhysRevB.87.165301}. Before we point out the most relevant for us results of those studies, we would like to mention the empirical in its nature studies (in the framework of self-consistent H\"uckel method and cluster model) showing that the Mn-atoms prefer to be substituted at the surface Ga-atoms instead of being adsorbed above them \cite{Surface.341.1995}. The deposition of Mn onto the GaAs surfaces happens mostly in the As-rich growth conditions. In this case, according to the Ref. \cite{PhysRevB.68.075202}, the Mn-atoms preferentially substitute Ga atoms in the cationic sublattice of GaAs bulk rather than build in the interstitial sites. The similar preference of the cation substitutional over interstitial position under As-rich conditions was predicted for isolated Mn-atoms on (001) GaAs surface in the DFT calculations \cite{PhysRevLett.89.227201}. 
Also in Ref \cite{PhysRevB.87.165301}, it was predicted in the DFT calculations that under As-rich conditions Mn-atoms are favorably incorporated into Ga sites at the $c(4\times4)$ reconstructed (001) GaAs surface with Mn coverage of $1/4$ mono-layer (ML). The DFT calculations in Ref. \cite{PhysRevB.87.165301} were motivated by the STM experiments reported therein, which demonstrated that reconstruction of Mn covered surface changes with the growth temperature. The experiments indicated also the existence of the $(2\times2)$ types of reconstructions. The Mn-induced surface reconstructions were also studied both by STM and DFT calculations for higher than $1/4$ Mn coverage, namely for Mn coverage of $1/2$ ML, $3/4$ ML, and 1 ML \cite{PhysRevB.69.121308}. The STM images revealed also the coexisting areas of $2\times2$ and $2\times4$ reconstructions on As-terminated (001) GaAs surface covered with Mn \cite{PhysRevB.69.121308}. The DFT calculations indicated that the $2\times4$ reconstructions should be preferential for lower Mn coverage \cite{PhysRevB.69.121308}. In spite of the fact that the experimental and theoretical studies of Mn covered (001) GaAs surfaces \cite{PhysRevLett.89.227201,PhysRevB.69.121308} were performed for higher Mn coverage than considered in this paper, they clearly indicate the importance of the $2\times4$ reconstructions and the fact that in As-rich conditions the Mn-atoms substitute Ga-atoms. Therefore, to get an overall understanding of the incorporation mechanisms of the isolated Mn-pairs on the (001) GaAs surface, and facilitate comparison between various surface geometries, we have performed extensive studies of all possible non-equivalent substitutional positions of the Mn-pairs onto (001) reconstructed GaAs surfaces: $(2\times1)$, $\beta(2\times4)$, $\beta2(2\times4)$, under As-rich conditions, employing the identical computational tool. In contrast to calculations reported in Refs. \cite{PhysRevLett.89.227201} and \cite{PhysRevB.69.121308}, where only the standard approximations to the DFT were employed, we decided to perform calculations also within L(S)DA+U procedure \cite{PhysRevB.57.1505,PhysRevB.70.235209}, which proved to lead to better agreement with experimental results for bulk DMSs \cite{RevModPhys.86.187,RevModPhys.86.855}, in addition to the standard L(S)DA approximation. Therefore, the present studies constitute complement of the previous ones \cite{PhysRevLett.89.227201,PhysRevB.69.121308} concerning the stability and morphology of the Mn covered reconstructed (001) GaAs surfaces and provide new knowledge of physical mechanisms that determine the magnetic interactions between two Mn-atoms at the (001) GaAs surfaces.

The magnetic interactions of Mn ions on GaAs surfaces of various orientations are only very weakly understood up to now. Strandberg \textit{et al.} \cite{PhysRevB.81.054401} studied the Mn-pairs at different crystallographic orientations at (110) GaAs surface by employing the Kinetic Tight Binding model. They showed that the long-range interactions were anisotropic in terms of orientations and distances between the Mn-pairs. They also demonstrated that the magnetic ions prefer to be ferromagnetically arranged. The anisotropic character of the effective exchange constant for the Mn-pairs differently oriented at (110) surface has been shown in Ref. \cite{PhysRevB.85.155306}, by  using the Density Functional method. In those two papers, the most common cleaved surface employed in cross sectional STM studies has been considered, albeit it is not the most common GaAs growth surface (which is the (001) one). In the literature, there is lack of information about the exchange interaction between the magnetic ions (particularly in dilute regime) placed onto the experimentally observed reconstructed (001) GaAs surfaces.

Detailed knowledge of the magnetic interactions at the reconstructed surfaces is essential for fabricating new high speed spintronics devices. Therefore, in this work we investigate magnetic properties of Mn-pairs incorporated into the differently reconstructed (001) GaAs surfaces in diluted case. Moreover, we report the role of the surface reconstruction in the energetics of the Mn incorporation process and stability of the Mn-GaAs surfaces. 

The main objection of the present paper is to study the stability, magnetic interactions, and electronic structure of isolated Mn-pair incorporated into reconstructed (001) GaAs surfaces. This should allow us also to deepen the understanding of the relation between the preferential distribution of the Mn-pairs incorporated into the GaAs and the uniaxial magnetic anisotropy, which has been recently demonstrated \cite{PhysRevLett.108.237203}. The separated Mn-pairs have been chosen, since it is known that Mn-atoms in GaAs have tendency to cluster \cite{PhysRevB.63.233205}, and there are some experimental suggestions that such Mn-pairs can really form in some growth conditions as described in Ref. \cite{PhysRevLett.108.237203}. We note that according to Ref. \cite{PhysRevB.65.201303} the Mn-ions forming pairs occupy Ga substitational positions, and once formed, they remain stable through the further growth process as well as during post-growth annealing at low temperatures $T_a<T_g$ which are employed to diffuse out Mn in interstitial positions. Of course, such effects as the existence of single Mn-atoms at the surface, interaction of the Mn-atoms with various types of surface structural defects, and/or disorder of Mn-pairs would correspond probably to more realistic situation to be encountered in a growth process, however, such studies seems to lie outside the scope of ab initio calculations at present. Nevertheless, we believe that our studies shed light on the mechanisms of stability and mutual magnetic interactions between Mn-atoms at surfaces. This has particular importance in times when individual atoms can be placed at surfaces employing direct techniques such as STM, for example.

The paper is organized as follows. In section 2, we present computational details. The results are presented and discussed in section 3. Here we deal with the morphology, energetics, magnetic interactions, and electronic structure of the Mn-pairs on the reconstructed (001) GaAs surfaces. Finally, the paper is concluded in section 4.  
\section{Computational details}
\label{details}
The calculations are performed within the DFT \cite{PhysRev.136.B864,PhysRev.140.A1133} computational scheme employing the L(S)DA+U approach and parameterization of the exchange-correlation functional provided by Ceperly and Adler (CA) \cite{PhysRevLett.45.566}, as it is implemented in SIESTA code \cite{PhysRevB.53.R10441}. The electron ion-core interactions are represented by norm-conserving pseudopotentials of the Troullier-Martins type \cite{PhysRevB.43.1993} with non-linear core corrections \cite{PhysRevB.26.1738}. The electron wave-functions are expanded into a flexible multiple centered atom basis set of numerical atomic orbitals \cite{StatSolb.215.1999}. In the calculations we use double-$\zeta$ for the $s$ and $p$ shells of any element and a triple-$\zeta$ basis set for the Mn $3d$ shell. The cutoff of $300 \, \mathrm{Ry}$ is used for the real space mesh. The Brillouin Zone (BZ) integration is performed by means of the k-grid parameter of 30 \AA{}, which corresponds to 16 k-points in the full Brillouin Zone on the $(4\times 4\times 2)$ shifted k-grid one. For the L(S)DA+U calculations, we adopted the value of the U parameter equal to 4.5 eV for Mn $3d$ states, which is in perfect agreement with the previous photoemission data \cite{PhysRevB.64.125304}, and consistent with previous L(S)DA+U calculations for the bulk \cite{RevModPhys.78.809,PhysRevB.70.235209}.

The important issue of large-scale computations is to evaluate their internal accuracy. In particular, we are interested in energetics of the surface calculations, therefore, the convergence of the surface free energy is systematically checked. We have essentially four parameters that determine the internal accuracy (or convergence) of the computations, i.e., the accuracy of computations for chosen density functional and pseudopotentials. These parameters are: (i) kinetic energy cutoff, or in the SIESTA code the cutoff for the real space, (ii) the number of k-points, or in the SIESTA code the so-called k-grid parameter, (iii) the number of layers in the slab, and (iv) width of the vacuum region in the supercell. The two first parameters have been defined above, and the other two will be determined later on, just describing the geometry of the systems. We have tested the convergence of the surface energy with respect to all four parameters. We increased the value of one parameter systematically, simultaneously keeping the other three ones constant, and observed the convergence of the surface energy. Within this procedure, we estimated the surface energy convergence error connected to the employed k-point grid to be of the order of 0.05 meV/\AA$^2$. Similarly, the chosen cutoff for the real space mesh leads to the estimated error in the surface free energy of the order of 0.07 meV/\AA$^2$. In similar manner we estimate the convergence errors in the surface free energy connected to the width of the vacuum layer and the number of layers in the slab. Altogether, the estimated internal accuracy of calculations of the surface free energy (being the sum of four errors) is not smaller than 0.37 meV/\AA$^2$, which is quite accurate, if one compares this value with the typical values of the (001) GaAs surface energies \cite{Murdick.17.2005}, lying in the range 40 to 100 meV/\AA$^2$.


\subsection{Model of a surface}
\begin{figure}[tp]
\centering
\includegraphics[width=0.95\columnwidth]{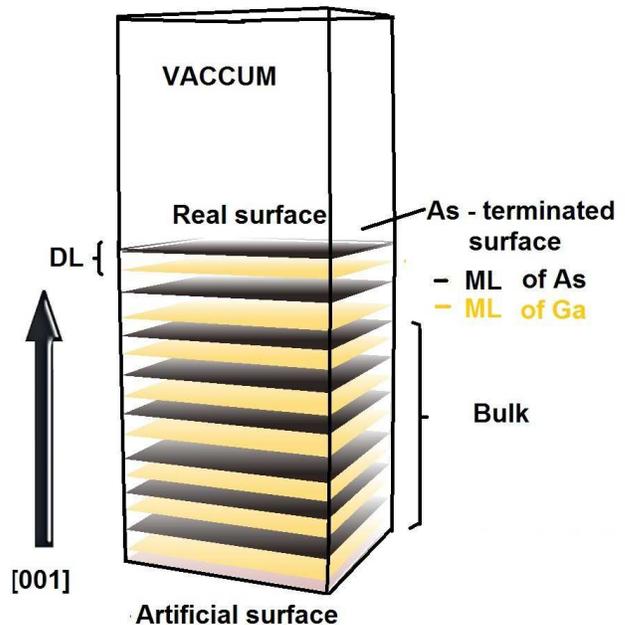}
\caption{\label{fig:model} The schematic diagram of the supercell used in the calculations.
The supercell consists of 8 double layers of As-Ga. Each monolayer contains 16 atoms. The artificial surface denotes the surface with hydrogen saturated dangling bonds, of Ga ML which mimics the bulk type of bindings.}
\end{figure}

In order to investigate  the physical properties of substitutional
isolated Mn-pairs on the GaAs(001) surfaces,
we used supercell geometry and construct the slab system as it is presented in Fig. \ref{fig:model}.
To model (001) GaAs surface, 8 double As-Ga layers (DLs) lying in the (001) crystallographic planes are used. The GaAs crystal is represented by a standard zinc blende (zb) cell with calculated lattice parameter equal to 5.639 \AA{}, which is in good agreement with experimental value of 5.648 \AA{} \cite{b1}. If the slab is not thick enough, the dangling bond states on the two sides of the slab might interact with each other and give rise to artificial charge transfer from top surface of the slab to its bottom. To avoid this effect and decouple the two sides of the slab, we saturate the dangling bonds from the bottom side of the slab by a monolayer of pseudo-Hydrogen atoms with fractional charge equal to $Z=1.25$. Each of the Ga-atom is saturated by two pseudo-atoms, to mimic the bulk types of bonds. In order to simulate independent crystal surface, 16 \AA{} of vacuum is added (it corresponds to 12 MLs of the bulk crystal), with the slab dipole correction option enabled, as it is implemented in SIESTA code \cite{PhysRevB.53.R10441}.
\begin{figure}
\centering
\includegraphics[width=1.0\columnwidth]{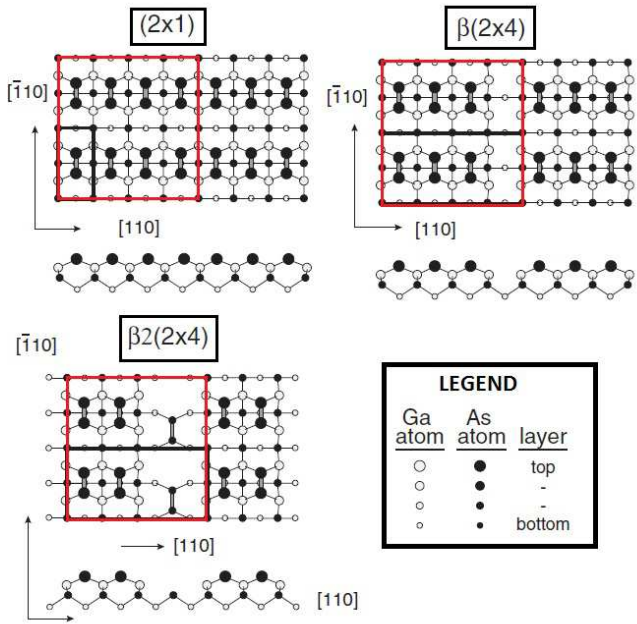}
\caption{\label{fig:sch}  The top-view and side-view schematic diagrams of (001) GaAs reconstructed surfaces under As-rich conditions used in this work. On the surface, As-As dimerization along [$\bar{1}$10] direction is clearly visible. Four monolayers from the top of the slab are shown. Positions in the uppermost atomic layers are indicated by larger symbols. The black and white balls indicate the arsenic and gallium atoms, respectively. The rectangular, red line denotes the lateral supercell chosen for the calculations.}
\end{figure}

We consider experimentally observed reconstructed (001) GaAs surfaces under As-rich conditions:
$\beta(2\times4)$ \cite{PhysRevLett.60.2176}, $\beta2(2\times4)$ \cite{ApplPhysLett.60.1992,Gallagher199231,Avery199591,PhysRevB.68.085321}, and also theoretically proposed $(2\times1)$ reconstruction \footnote{Experimentally, this reconstruction is observed on Si(001) surfaces} \cite{Colonna.109.2011}, which are presented in Fig. \ref{fig:sch}. All of the atoms in pure surfaces are fully relaxed until the maximal force on each atom reaches the value of 0.02 eV/\AA{}. These surfaces are our starting points for further calculations.

 To investigate the physical properties of isolated Mn-pairs (it models properly diluted case), we substitute two Ga-atoms by two Mn-atoms at the top most monolayer of Ga-atoms at pure (001) GaAs reconstructed surfaces. Our simulated supercells contain 288 atoms for $(2\times1)$, 284 atoms for $\beta(2\times4)$, and 276 atoms for $\beta2(2\times4)$ reconstructed surface. Each of the monolayer comprises 16 atoms, making the Mn-coverage of the layer equal to 1/8.  The surface area corresponds to the lateral ($4\times4$) cell of the dimensions 16\AA{}$\times$16\AA{}. During the optimization procedure, six MLs from the top of the slab are fully relaxed, whereas the bottom of the slab is fixed to reflect the bulk character of this part of the supercell.

\section{Results}
\label{results}

Here we present main results of our studies. First, we describe the energetic issues in subsection 3.1, and then we turn to the electronic and magnetic properties of the surfaces with incorporated Mn-pairs in subsection 3.2.

\subsection{Energetics}
\label{energetics}
In this section, we focus on structural and energetic properties from the standpoint of the different possible incorporations of Mn-pairs which are embedded onto cationic sublattice into the three reconstructed (001) GaAs surfaces: $(2\times1)$, $\beta(2\times4)$, and $\beta2(2\times4)$.

We discuss first each of the reconstructed surface, and then provide a comparison between the three studied reconstructions.

\subsubsection{Reconstruction \texorpdfstring{$(2 \times 1)$}{(2 x 1)}}

\begin{figure*}
\centering
\includegraphics[width=0.9\textwidth]{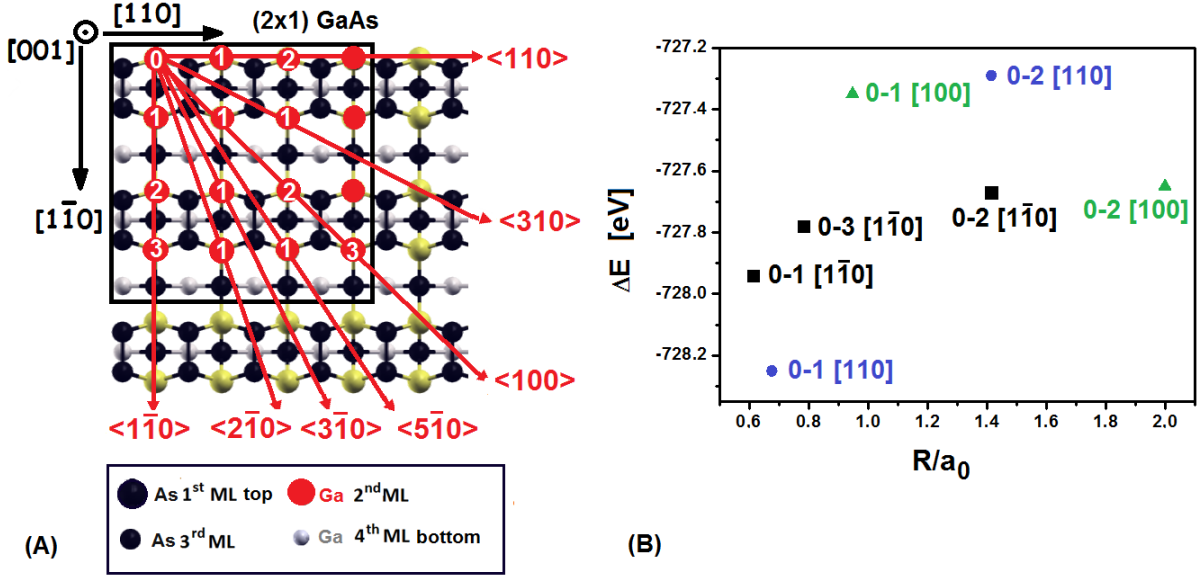}
\caption{\label{fig:recon2x1pion} (A) All non-equivalent positions of Mn-pair on the $(2\times1)$ reconstructed (001) surface. The two Mn-atoms, replacing Ga-atoms (red spheres), are marked 0 (first Mn-atom) and 1,2,3 (second Mn-atom), e.g. 0-3 [1$\bar{1}$0] denotes that the Mn-pair is along [1$\bar{1}$0] crystallographic direction, where Mn-atoms constituting the pair sit on 0 and 3 position as is depicted on panel (A). Note that the position 0-3 [1$\bar{1}$0] is equivalent to position 1-2 [1$\bar{1}$0] in a supercell, and so forth. All equivalent symmetry directions $<...>$ have been chosen in direct cubic coordinates [...], in such a way that all Mn-pairs appear in (001) plane. Positions of the atoms in the uppermost monolayers are indicated by larger symbols. Positions of the As-atoms are denoted with big black dots. Small grey dots indicate the positions of the Ga-atoms. (B) The stability of systems with Mn-pair incorporated into the $(2\times1)$ reconstructed surface, measured as the LDA energy differences  $\Delta E= E^{\textit{slab,Mn-pair}}_{tot}(MS) - E^{slab,pure}_{tot}$ as the function of the Mn-Mn distance. For clarity, the energetically less stable configurations (defined in panel A) have been omitted here. The two most stable configurations are the ones with Mn-atoms being nearest neighbors along [110] and [1$\bar{1}$0] directions.}
\end{figure*}

We start the discussion of energetics of the Mn-pair incorporated at the (001) GaAs surfaces with the case of the ($2\times1$) surface reconstruction. We treat this case as a prototype and describe the methodology used to study Mn-pairs on other reconstructed surfaces. Two Mn-atoms can be substituted on Ga sites of the $4\times4$ lateral unit cell (see Fig. \ref{fig:sch}) in many ways as it is indicated in Fig. \ref{fig:recon2x1pion}. We consider substitution of Mn-atoms into the second from top cationic (Ga) layer (the top layer consists of As atoms), i.e., for all considered configurations the Mn-pairs lie in the (001) crystallographic plane. First Mn-atom is substituted on Ga site indicated as "0", Mn(0). The possible position of the second Mn-atom is indicated by the crystallographic direction along which the pair can be placed $[kl0]$ and the integer $n$ $(n = 1,2,3,...)$ that indexes possible positions of the second Mn along this direction. The configuration of the Mn($0$)-Mn($n$)-pair along the $[kl0]$ direction will be indicated as $0-n[kl0]$ from now on. As it is seen in Fig. \ref{fig:recon2x1pion}, there are three possible configurations of the Mn-pair along [1$\bar{1}$0] direction, but only two along [110] and [100] directions, and only one Mn-pair configuration along the [310], [5$\bar{1}$0], [3$\bar{1}$0], and [2$\bar{1}$0] directions.  Because of the periodic boundary conditions imposed on the lateral supercell, some of the indicated possible Mn positions are equivalent; specifically, the Mn(0)-Mn(3) pair is equivalent to the Mn(1)-Mn(2) pair. Altogether, one has 11 nonequivalent Mn-pairs in the $4\times4$ lateral unit cell of the ($2\times1$) reconstructed surface: three with Mn-atoms being nearest neighbors on cationic sublattice (indicated as $0-1[1\bar{1}0]$, $0-1[110]$, $0-3[1\bar{1}0])$ with the Mn-Mn distances R = 3.46\AA{}, R = 3.81\AA{}, and R = 4.41\AA{}, respectively; two with Mn-atoms being the second nearest neighbors ($0-1[100]$, $0-1[2\bar{1}0])$, R = 5.34\AA{} and R= 5.95\AA{}, respectively; two with Mn-atoms being the third nearest neighbors ($0-2[1\bar{1}
0]$, $0-2[110]$) with R = 7.98\AA{}; three with Mn-atoms being the fourth nearest neighbors ($0-1[310]$, $0-1[3\bar{1}0]$, $0-1[5\bar{1}0]$), R = 8.73\AA{}, R = 8.92\AA{}, R = 9.12\AA{}, respectively; and one with Mn-atoms being the fifth nearest neighbors ($0-2[100]$), R = 11.29\AA{}.

Having defined 11 configurations of the Mn-pairs placed onto the $4\times4$ lateral unit cell, we are now in the position to determine their relative energetic stability.
We define the incorporation energy of Mn-pair at the surface in configuration $0-n[kl0]$ employing the standard expression \cite{b2}:

\begin{eqnarray}
 \lefteqn{E_{incorp}(0-n[k l 0]) \defeq E^{\textit{slab,Mn-pair}}_{tot}(0-n[k l 0])} \qquad \qquad \qquad \label{ads} \\
 & &  {} - E^{\textit{slab,pure}}_{tot} - (\mu_{Mn} - \mu_{Ga}) N_{Mn}, \nonumber
\end{eqnarray}
where $N_{Mn}$ is the number of Mn-ions substituted on the Ga sites ($N_{Mn}$ is always 2 in our studies), and $\mu_{Mn}$, $\mu_{Ga}$ are the chemical potentials of Mn- and Ga-ions, respectively.
However, since the number of substituted atoms in each considered configuration is identical, it is sufficient to consider only the total energy of the slab with $0-n[kl0]$ Mn-pair normalized to the total energy of the pure slab:
 \begin{eqnarray}
\label{diff}
 \Delta E(0-n[k l 0]) & \defeq & E^{\textit{slab,Mn-pair}}_{tot}(0-n[k l 0]) \\
  & & {} - E^{\textit{slab,pure}}_{tot}. \nonumber
\end{eqnarray}

The values of this energy for some of the considered configurations $0-n[kl0]$ versus the distance between the Mn-atoms are depicted in Fig. \ref{fig:recon2x1pion}(B).
It is clearly seen that the energetically most stable configuration of the Mn-pair on the $(2\times1)$ reconstructed (001) GaAs surface is the configuration with Mn-pair placed along the [110] direction with Mn-atoms being the first neighbors on the cationic sublattice. It is so, even then, the distance between Mn-atoms in the $0-1[1\bar{1}0]$ configuration is smaller than in the $0-1[110]$ one. It demonstrates directional preference of accommodating Mn-pair on the $(2\times1)$ reconstructed surface, which results from the local environment of the Mn-pair.   However, as expected and seen in Fig. 3(B), the configurations with Mn-atoms being further apart are generally energetically less favorable. It is strong indication of the tendency of Mn-atoms deposited on the surface to form pairs.

The energy difference between $0-1[110]$ and $0-1[1\bar{1}0]$ configurations of Mn-NN-pair is 0.4 eV per supercell with 288 atoms. It is obvious that the local changes of the geometry contribute considerably to the relative stability of these two configurations. It is worth to point out now that when we place Mn-NN-pair along [110] crystallographic direction, the symmetry is lowered to C$_v$ in comparison to pure ($2\times1)$ reconstructed (001) GaAs surface (where it is C$_{2v}$), whereas Mn-NN-pair placed along the $[1\bar{1}0]$ direction doesn't change the symmetry of the layer. As a consequence, we have noticed stronger relaxation of the atoms around the Mn-NN-pair along [110] direction, than for the Mn-pair along the $[1\bar{1}0]$ direction.
\begin{figure}
\centering
\includegraphics[width=0.8\columnwidth]{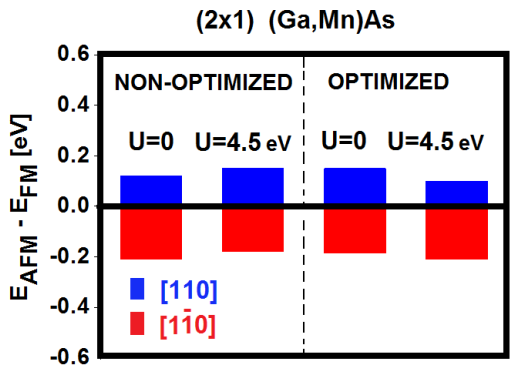}
\caption{\label{fig:SCFOPT2x1} The energy difference between AFM and FM Mn-spin alignment E$_{AFM}$-E$_{FM}$ in eV/supercell, for the nearest neighbor 0-1 Mn-pair along the [1$\bar{1}$0](red) and [110](blue) directions at $(2\times1)$ reconstructed surface, as calculated within standard L(S)DA (indicated by U=0) and L(S)DA+U (indicated by the used U=4.5eV value) methods. For Mn-NN-pair along [1$\bar{1}$0] direction, the magnetic ground state is anti-ferromagnetic (AFM), whereas for [110] direction, it is ferromagnetic one (FM). It is seen that optimization of atomic positions does not change magnetic state of the system.}
\end{figure}

All the results concerning relative stability of Mn-pairs described above have been obtained employing standard nonmagnetic calculations with LDA exchange-correlation functional. It is apparent that the change of the energy owing to the local rearrangement of atoms in the neighborhood of Mn-pair is roughly of the order of the magnetic energy of the Mn-pair. Therefore, we included the magnetic interaction into our study employing the standard local spin density approximation (L(S)DA) and L(S)DA+U calculations. Employing these two methods, we consider parallel (ferromagnetic - FM) and antiparallel (anti-ferromagnetic - AFM) ordering of localized magnetic moments of Mn-atoms constituting the Mn-pair on the nearest neighbor cationic positions along [110] and $[1\bar{1}0]$ directions. We use the standard value of U = 4.5 eV parameter for Mn-atom that has been routinely used in many calculations involving Mn \cite{PhysRevB.64.125304,RevModPhys.78.809,PhysRevB.70.235209}.
For non-optimized atomic positions\footnote{"Non-optimized atomic positions" refer to the ideal position of GaAs surface atoms. In other words, we have substituted the Ga-atoms by the Mn-atoms without allowing the atoms to relax.} around the Mn-pair, the AFM ordering of Mn magnetic moments in Mn-pair is more favorable than FM one for $[1\bar{1}0]$ direction, whereas for Mn-pair along the [110] direction the relation is reversed. The full relaxation of the slab does not change this picture as it is illustrated in Fig. \ref{fig:SCFOPT2x1}. Qualitatively the same picture is obtained for both L(S)DA and L(S)DA+U functionals. From the point of view of magnetic interaction between the Mn-atoms, the chemical arrangement of atoms around the Mn-pair plays the crucial role, whereas the small changes of atomic geometry resulting from the relaxation of atomic positions has a negligible influence on it. In other words, only the chemical arrangement is able to change the direction of the magnetic moments on the Mn-atoms, and hence the magnetic state (FM or AFM) of the (Ga,Mn)As system. For the L(S)DA+U method, the energies defined in equation \ref{diff} are as follows $\Delta$E($0-1[110]$; FM) = $-735.90$ eV; $\Delta$E($0-1[110]$; AFM) = $-735.81$ eV; $\Delta$E($0-1[1\bar{1}0]$; AFM) = $-735.54$; $\Delta$E($0-1[1\bar{1}0]$; FM) = $-735.34$ eV. Therefore, the most stable configuration of the Mn-pair incorporated into the $(2\times1)$ reconstructed (001) GaAs surface is $0-1[110]$ with parallel magnetic moments of two Mn-atoms.

The discussion for the $(2\times1)$ reconstructed surface sheds light on the energetics of the Mn-pair incorporation at this surface. It clearly demonstrates the interplay between the magnetic interactions of the Mn-atoms and atomic relaxations around them. Further, we follow the discussion for Mn-pairs incorporated into $\beta(2\times4)$ and $\beta2(2\times4)$ reconstructed surfaces. However, in the light of results obtained for $(2\times1)$ reconstructed surface, we focus the discussion on Mn-NN-pairs placed along [110] and $[1\bar{1}0]$ crystallographic directions.

\subsubsection{Reconstruction \texorpdfstring{$\beta(2\times4)$}{beta(2 x 4)}}

\begin{figure}
\centering
\includegraphics[width=0.9\columnwidth]{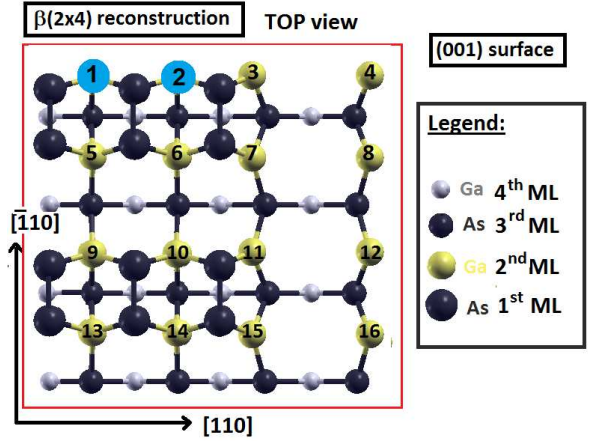}
\caption{\label{fig:r2x4st} Positions of Mn-pairs at the $\boldsymbol{\beta(2\times4)}$ reconstructed surface. The energetically most favorable position is for Mn-pair indicated as 1 \& 2 along [110] direction with FM ordering of spins. For emphasis, this pair is indicated by the blue color. This is true for both L(S)DA and L(S)DA+U calculations. All other cationic positions, where Mn-atoms can be substituted are indicated by yellow color and numbered 3-16. Note that 5 \& 6 positions of Mn-atoms are equivalent to 1 \& 2 and lead to the same energy. Positions of the As-atoms are indicated with big and small black dots  for the highest and deeper As-layers, respectively.}
\end{figure}

We start the discussion with a survey of possible placements of the Mn-NN-pairs on the $\beta(2\times4)$ (001) GaAs surface (see Fig. \ref{fig:r2x4st}).
In the chosen supercell of $C_{2v}$ symmetry, the [110] and [1$\bar{1}$0] directions are non-equivalent. Along each of these directions, we can place the nearest neighbor
Mn-pair on four equivalent ways on the cationic sublattice. They are, 1 \& 2, 5 \& 6, 9 \& 10, 13 \& 14, for [110], and 1 \& 5, 2 \& 6, 9 \& 13, 10 \& 14 for [1$\bar{1}$0] directions, respectively.

However, along the [110] and [1$\bar{1}$0] directions we can place Mn nearest neighbor pairs also in non-equivalent positions, such as 1 \& 2, 2 \& 3, 3 \& 4, along [110], and 6 \& 10, 7 \& 11, 10 \& 14, 11 \& 15 along [1$\bar{1}$0]. The difference in energy for the nearest neighbor (NN) configurations mostly stems from the fact that the surroundings of Mn-NN-pairs are different, but also to a lesser extent, that the distance (after relaxation of atoms) between the Mn-atoms constituting the pair also has been changed.

From Fig. \ref{fig:r2x4st}, one can also deduce the positions of the Mn-pairs with longer than nearest neighbor distances between Mn-atoms. We have performed calculations for all non-equivalent positions of the Mn-pairs. It turns out that the pairs with Mn-atoms being the nearest neighbors are energetically preferable in comparison to the configurations with larger distances between the Mn-atoms.

However, the relative energy differences among this class of configurations are dependent on the
 alignment of magnetic moments of the Mn-atoms (FM or AFM).
This is illustrated in Table \ref{tab:nonAll}, where the difference of the total energies of
 slab with Mn-NN-pair and the pure slab (i.e., without Mn-NN-pair) are given for the L(S)DA and L(S)DA+U method.

\begin{table}\footnotesize
\caption{The stability of the systems with the incorporated Mn-NN-pairs at the $\beta(2\times4)$ reconstructed (001) GaAs surface measured by the energy  $\Delta E= E^{slab,Mn}_{tot}(MS) - E^{slab,pure}_{tot}$, given in eV per supercell. The Mn-pair positions are explained in Fig. \ref{fig:r2x4st}. $E^{slab,pure}_{tot}=-39813.12$ eV per supercell with 284 atoms. The parallel and antiparallel spin alignment of the Mn-NN-pair is indicated by FM and AFM, respectively. The results are obtained employing standard L(S)DA and L(S)DA+U ($U = 4.5$ eV) approaches.} 
   \def\arraystretch{1.5}
\label{tab:nonAll}
\begin{center}
  \begin{tabular}{ | c|  c|  c|  c| c |}
    \hline
 $\beta(2\times4)$ &   \multicolumn{2}{|c|}{$\Delta E_{L(S)DA} $ [eV/cell]} &   \multicolumn{2}{|c|}{$\Delta E_{L(S)DA+U}$ [eV/cell]}  \\
\hline
Mn-NN-pair position & FM & AFM & FM & AFM \\
\hline
   [110] 1 \& 2&  -737.743&-737.448 & -743.792& -743.555\\
\hline
    [110] 2 \& 3&  -737.582&-737.456& -743.765& -743.666\\
\hline
    [110] 3 \& 4 &  -737.561 &-737.413  & -743.756 & -743.636\\
\hline
\hline
   [1$\bar{1}$0] 6 \& 10  & -737.415& -737.455& -743.521 &-743.582\\
\hline
    [1$\bar{1}$0] 7 \& 11  & -737.095& -737.053& -743.443& -743.430\\
\hline
   [1$\bar{1}$0] 10 \& 14  & -737.539& -737.557& -743.627& -743.623\\
\hline
    [1$\bar{1}$0] 11 \& 15&   -737.195 & -737.318&-743.605& -743.697\\
\hline
  \end{tabular}
\end{center}
\end{table}

Note that the total energy of the pure slab can be considered as a reference energy. This energy is identical for the L(S)DA and L(S)DA+U case,
since for the atoms in the pure slab (As, Ga, \textsl{H}) the Hubbard term $U$ has been always taken as zero.

The energies of all non-equivalent Mn-pair arrangements (with Mn-atoms being the nearest neighbors) are presented in Table \ref{tab:nonAll}.

Among the Mn-NN-pairs placed along [110] direction, the lowest total energy is for FM Mn-spin arrangements and the Mn-NN-pairs
1 \& 2 (see Fig. \ref{fig:r2x4st}, where this pair is indicated in blue). This is true for both L(S)DA and L(S)DA+U approaches. Among all Mn-NN-pairs placed along [1$\bar{1}$0] direction, the lowest total energy is for the AFM spin configuration for the Mn-NN-pair 10 \& 14 in L(S)DA approach (see Fig. \ref{fig:r2x4st}) and for the Mn-NN-pair 11 \& 15 when L(S)DA+U approach is employed. Generally, the L(S)DA and L(S)DA+U approaches lead to identical trends concerning the interplay between magnetic ordering and local environment of Mn-NN-pairs. For all Mn-NN-pairs along [110] direction, the FM alignment  of Mn-spins is favorable over the AFM one. For Mn-NN-pairs along [1$\bar{1}$0] directions,  generally the AFM spin alignment  leads to lower energies than the FM one, however  it is not the case for Mn-NN-pair numbered 7 \& 11.

\subsubsection{Reconstruction \texorpdfstring{$\beta2(2 \times 4)$}{beta2(2 x 4)}}

We discuss now structural and energetic properties of the Mn-NN-pairs on cationic sublattice and placed onto the
$\beta2(2\times4)$ reconstructed (001) surface.
\begin{figure}
\includegraphics[width=0.9\columnwidth]{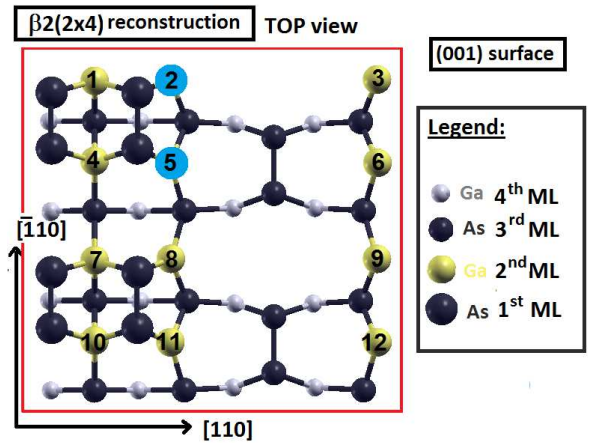}
\caption{\label{fig:r22x4st} Positions of Mn-pairs at the $\boldsymbol{\beta2(2\times4)}$ reconstructed surface. According to L(S)DA+U approach, the energetically most favorable configuration is the one with Mn-NN-pair (indicated as 2 \& 5) along [1$\bar{1}$0] direction with AFM ordering of spins. This configuration has been indicated by blue color. All other possible position of Mn-atoms on cationic sublattice are indicated by yellow color and numbered 3-12.  Positions of the As-atoms are indicated with big and small black dots  for the highest and deeper As-layers, respectively.}
\end{figure}
As in the $\beta(2\times4)$ case, we start the discussion with a survey of possible placements of the Mn-NN-pairs
onto the $\beta2(2\times4)$ (001) GaAs surface (see Fig. \ref{fig:r22x4st}).
In the chosen symmetry of the supercell, the [110] and [1$\bar{1}$0] directions are non-equivalent.
 Along the [110] crystallographic direction, one can place the nearest neighbor
Mn-pair in eight equivalent ways on the cationic sublattice, 1 \& 2, 4 \& 5, 7 \& 8, 10 \& 11, 1 \& 3, 4 \& 6, 7 \& 9, and 10 \& 12, which results in one class of non-equivalent positions represented by 1 \& 2. Along the [1$\bar{1}$0] direction, there are four classes of non-equivalent positions of the Mn-NN-pair. They were chosen to be 1 \& 4, 2 \& 5, 4 \& 7, and 5 \& 8. Note that 1 \& 4 and 7 \& 10 or 5 \& 8 and 6 \& 9 are equivalent.
Our results show that the pairs with Mn-atoms
 being the nearest neighbors are energetically preferable in comparison to the configurations with larger distances between the Mn-atoms.
In table \ref{tab:nonAll1}, we present the relative energy differences among all nearest neighbor classes of Mn-pair configurations for FM and AFM alignments of magnetic moments of the Mn-NN-pair.

The analysis of energies presented in table \ref{tab:nonAll1} reveals following picture. For Mn-NN-pair along [110] direction, the FM alignment of spins is energetically more favorable than the AFM one (in both L(S)DA and L(S)DA+U approaches), as it was observed for the $\beta(2\times4)$ reconstruction.  However this configuration (in contrast to the case of the $\beta(2\times4)$ reconstruction) is not the most stable one.
The most stable configurations are observed for Mn-NN-pairs along [1$\bar{1}0$] directions. According to the standard L(S)DA the most stable configuration is the one with Mn-atoms 4 \& 7 along [1$\bar{1}$0] with AFM Mn-spins alignment, whereas the L(S)DA+U predicts the 2 \& 5 pair (indicated by blue color in Fig. \ref{fig:r22x4st}) with AFM spin alignment to be the most energetically favorable. However, one has to have in mind that the differences in energies of different configurations are extremely tiny.

\begin{table}\footnotesize
\caption{The stability of the systems with the incorporated Mn-NN-pairs at the $\beta2(2\times4)$ reconstructed (001) GaAs surface measured by the energy  $\Delta E= E^{slab,Mn}_{tot}(MS) - E^{slab,pure}_{tot}$, given in eV per supercell. The Mn-pair positions are explained in Fig. \ref{fig:r22x4st}. $E^{slab,pure}_{tot}=-38570.17$ eV per supercell with 274 atoms. The parallel and antiparallel spin alignment of the Mn-NN-pair is indicated by FM and AFM, respectively. The results are obtained employing standard L(S)DA and L(S)DA+U ($U = 4.5$ eV) approaches.} 
   \def\arraystretch{1.5}
\label{tab:nonAll1}
\begin{center}
  \begin{tabular}{ | c | c | c | c | c |}
    \hline
$\beta2(2\times4)$ &   \multicolumn{2}{|c|}{$\Delta E_{L(S)DA} $ [eV/cell]} &   \multicolumn{2}{|c|}{$\Delta E_{L(S)DA+U}$ [eV/cell]}  \\
\hline
Mn-NN-pair position & FM & AFM & FM & AFM \\
\hline
   [110] (1 \& 2)&  -733.684&-733.483 & -739.621& -739.465\\
\hline
\hline
    [1$\bar{1}$0] (1 \& 4)  &-733.825 & -733.852& -739.601& -739.604\\
\hline
    [1$\bar{1}$0] (2 \& 5)&   -733.531&-733.571&-739.610&  -739.693\\
\hline
   [1$\bar{1}$0] (4 \& 7)  & -733.855 &-733.892&-739.614&-739.552\\
\hline
    [1$\bar{1}$0] (5 \& 8)  & -733.483& -733.386& -739.530& -739.459\\

\hline
  \end{tabular}
\end{center}
\end{table}

\subsubsection{Comparison between different reconstructions}
In this section, we make an attempt to generalize our theoretical studies presented in the previous sections and get physical understanding of the structural and energetic properties of (001) GaAs surfaces incorporated with Mn-pairs.

Our results show that the energy of the system strongly depends on the position, orientation, the distance between the Mn-atoms, and relative alignment of Mn-spins. The Mn-pairs prefer to occupy the nearest neighbor positions (NN) independently on the reconstruction type at the surface.
\begin{figure}
\centering
\includegraphics[width=0.95\columnwidth]{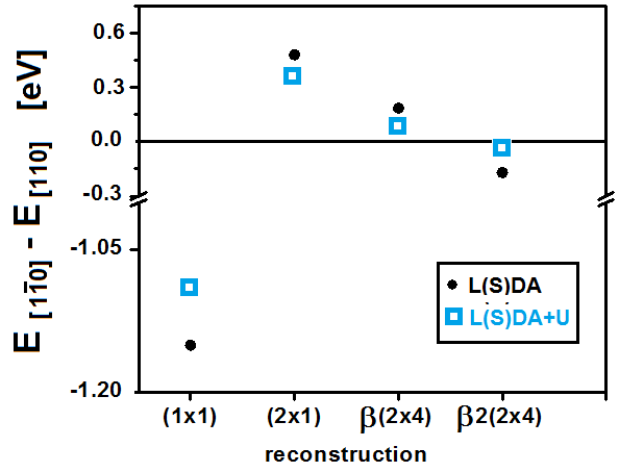}
\caption{\label{fig:compare} Comparison of the most stable configurations of the Mn-pairs at surfaces with various reconstructions. The energy difference between the most stable NN configuration at [1$\bar{1}$0] and [110] crystallographic directions, as obtained within the L(S)DA and L(S)DA+U schemes.}
\end{figure}

It turns out that for Mn-NN-pairs there are two crystallographic orientations leading to the energetically most stable configurations, namely the [110] and [1$\bar{1}$0] crystallographic directions. The energy differences between the most stable configurations along the [1$\bar{1}$0] and [110] directions (as obtained within the L(S)DA and L(S)DA+U schemes) are summarized for all considered surface reconstructions in Fig. \ref{fig:compare}. One can see that the energetically preferential orientation of the Mn-NN-pair depends on the surface reconstructions. In other words, it is plausible that during the growth process the Mn-ions would accommodate positions along the energetically preferential directions. The very recent measurements have shown that there exist magnetic inhomogeneities on sub-millimeter length scales in (Ga,Mn)As samples \cite{PhysRevLett.93.177201}, which can be assigned to anisotropic distribution of Mn-atoms arrangements. 

One of the important objective of this study has been the investigation of the relation between the preferential distribution of the Mn-pairs incorporated into the (001) GaAs surfaces and the origin of the uniaxial magnetic anisotropy in the (Ga,Mn)As samples, which, as it was shown recently by the authors \cite{PhysRevLett.108.237203}, can be ascribed to the breaking cubic symmetry of the bulk crystals by inhomogeneous distribution of Mn-atoms \cite{PhysRevLett.108.237203}. Here we have shown that the energetically preferential positions of the Mn-pairs depend on the surface reconstruction, and therefore, in principle on the growth conditions. However, our static calculations cannot provide a direct hint how the growth process leads to the predicted morphology of (001) GaAs surfaces with incorporated Mn-pairs. To reach this goal Molecular Dynamics or Kinetic Monte Carlo studies would be desirable, and this could be challenging aim of the future research. Nevertheless, our studies corroborate directly the existence of the preferential distribution of Mn-pairs at the most important reconstructed (001) GaAs surfaces and also indirectly the plausibility that various distributions can be achieved by suitable growth conditions.

Finally we compare the stability of the pure surface with the stability of the just calculated surfaces with the incorporated {Mn-pairs}, which could mimic the surfaces of (Ga,Mn)As. Stability of the structure can be determined by the standard thermodynamics, and have been well established for systems in equilibrium \cite{Qian}. The most stable surface structure is determined by the minimum of the surface free energy $\gamma$, which is defined for a slab by the equation below \cite{b2, TSS} :
\begin{equation}
\label{free}
\gamma =\frac{1}{A}(E_{surf}-\sum_{i}\mu_{i}N_{i}),
\end{equation}
where $A$ is the surface area of the slab within the supercell, $\mu_i$ is the chemical potential of species \textit{i},  and $E_{surf}$ is the surface energy defined as:
\begin{equation}
\label{free1}
E_{surf}=\frac{1}{2}(E_{slab}-n E_{bulk}),
\end{equation}
 where $E_{slab}$ denotes the total energy of the supercell, $n$ is the number of bulk unit cells contained in the slab cell, and having energy $E_{bulk}$. In order to plot the surface free energy versus the thermodynamically allowed range of chemical potential, the surface free energy can be re-expressed as a function of the difference in the chemical potentials of the atomic As and As-bulk, $\Delta\mu=\mu_{As}-\mu_{As}^{bulk}$. For the thermal equilibrium conditions \cite{Qian,TSS}, one obtains:
\begin{equation}
\label{G2}
\begin{split}
\gamma &=\frac{1}{A}( E_{surf}-N_{Ga}E_{GaAs}^{bulk}-N_{Mn}E_{MnAs}^{bulk}\\
&-(\mu_{As}-\mu_{As}^{bulk})\Delta N-\mu_{As}^{bulk}\Delta N )  ,
\end{split}
\end{equation}
where the stoichiometry parameter is defined as $\Delta N=(N_{As}-N_{Ga}-N_{Mn})$, and $E_{GaAs}^{bulk}$ and $E_{MnAs}^{bulk}$ are  the cohesive energies of GaAs and MnAs, respectively, when $T\rightarrow0K$. Our calculated heat of formation for MnAs is $\Delta H_{f}^{MnAs}=-0.52 $ eV, and for GaAs is $\Delta H_{f}^{GaAs}=-0.73$ eV per unit cell with 2 atoms for both systems.
The stoichiometry parameter $\Delta N$ determines the slope of the surface energy versus the difference in chemical potential ($\mu_{As}-\mu_{As}^{bulk}$). The stoichiometry parameter is defined by applying the method of Chetty and Martin \cite{PhysRevB.44.5568}, which utilizes the symmetry of the crystals and is commonly used approach. For example, following counting method of Ref. \cite{PhysRevB.44.5568}, $\Delta N$ for the ideal $(1\times1)$ As-terminated surface is equal to $\frac{1}{2}$. In order to understand this counting rule one can think of a symmetric slab with two identical, As-terminated surfaces. This slab has one As-atom more than Ga across the slab, so there is $\frac{1}{2}$ additional As-atom per $(1\times1)$ surface unit cell \cite{TSS}. By this procedure, one obtains stoichiometry parameter $\Delta N = \frac{1}{2}$ for $(1\times1)$, and $(2\times1)$ surfaces, and $\Delta N = \frac{1}{4}$ for $\beta(2\times4)$, and $\beta2(2\times4)$, surfaces, where $\Delta N$ is counted per $(1\times1)$ lateral unit cell. Note that the lateral unit cell employed in the present studies is 16 times larger than the $(1\times1)$ unit cell, and therefore, the stoichiometry parameter takes the values: $\Delta N = 8$ for $(2\times1)$, and  $\Delta N = 4$ for $\beta(2\times4)$, and $\beta2(2\times4)$ surface reconstructions.

\begin{figure}
\includegraphics[width=0.95\columnwidth]{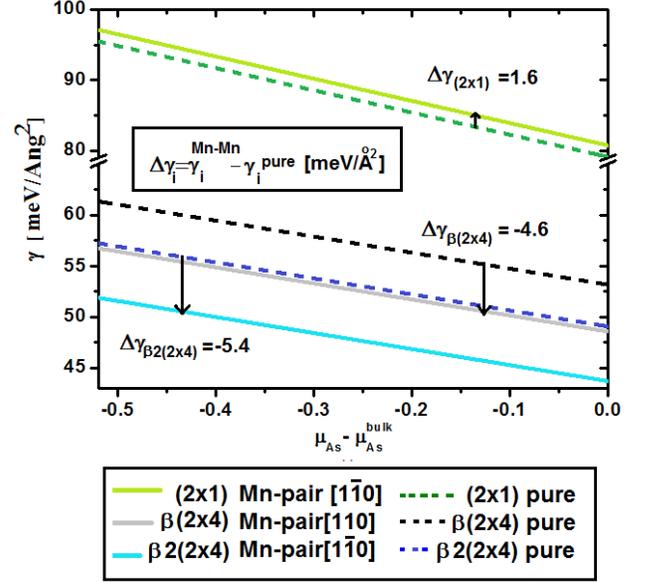}
\caption{\label{fig:stability}  Surface free energy diagrams for pure (001) GaAs, As-terminated surface reconstructions: $(2\times1)$, $\beta(2\times4)$,
$\beta2(2\times4)$ (denoted by the dashed lines), and for the most energetically preferable Mn-NN-pair substitutional incorporations onto these reconstructed surfaces (denoted by the straight lines) as predicted in this paper. Thermodynamically allowed range of the difference of chemical potentials of As-atom and As-bulk is between $-0.52$ eV (Ga-rich conditions) and $0.00$ eV (As-rich conditions). The arrows indicate the change in the surface free energy $\Delta\gamma_i$ owing to the incorporation of Mn-pair.}
\end{figure}

The surface energies for the systems studied in this paper are presented in Fig. \ref{fig:stability} for L(S)DA approach. There, one can see that the pure $\beta2(2\times4)$ reconstructed (001) GaAs surface is the most stable one for the whole range of chemical potential. This is a consequence of the dimerization of the atoms at the surface, which reduces the numbers of dangling bonds and creates the $sp^3$ like bonds. The difference between the $\beta(2\times4)$ and $\beta2(2\times4)$ surface energies is $4$ meV/\AA$^{2}$, which is the same order of magnitude as reported previously in literature, where this difference was determined to lie in the range of $2-3$ meV/\AA$^{2}$\cite{PhysRevLett.85.3890,PhysRevLett.89.227201}. Moreover, one can see that all the reconstructions considered in Fig. \ref{fig:stability} become more stable at the As-rich limit, i.e., $\mu_{As}-\mu_{As}^{bulk}$ approaching zero. The $\beta2(2\times4)$ reconstructed surface with Mn-NN-pair along [1$\bar{1}$0] direction is found to be energetically the most favorable of all Mn-reconstructed surfaces considered in this paper. The incorporation of the Mn-NN-pair into substitutional position for $\beta(2\times4)$ and $\beta2(2\times4)$ reconstructed surfaces stabilizes this surface by 5.4 meV/\AA$^2$, and 4.6 meV/\AA$^2$, respectively, whereas in the case of the $(2\times1)$ surface reconstruction, the effect is reversed and the surface is destabilized by the 1.6 meV/\AA$^2$.

\subsection{Magnetic and Electronic Properties}
\label{magnetic}
In this section, we focus on the strength of the magnetic interaction of Mn-NN-pairs at studied surfaces. Then we examine the electronic structure and spin magnetic moments.

\begin{figure}
\includegraphics[width=0.95\columnwidth]{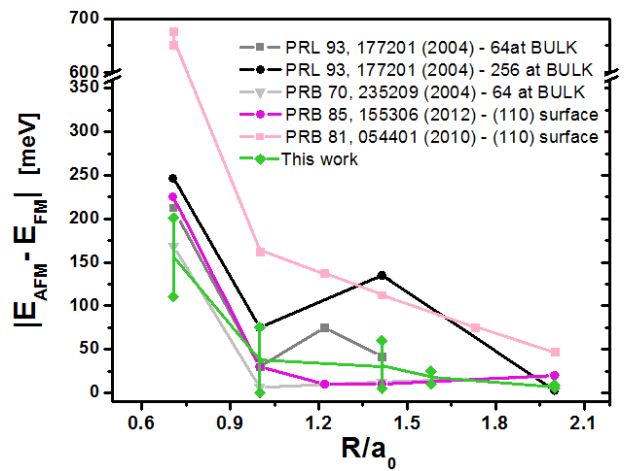}
\caption{\label{fig:recon}
The absolute value of the total energy difference between ferromagnetic and anti-ferromagnetic alignments of two Mn-spins as a function of Mn-Mn separation (in lattice constant units $a_0$) for the surface and bulk calculations. The green line presents the result of this work for the surface reconstruction $(2\times1)$, and the green vertical lines indicate the obtained range of the magnetic energies for various nonequivalent Mn-pair configurations separation between the Mn-ions. The internal panel shows the sources of data.}
\end{figure}

In order to show how the strength of the magnetic interaction of the Mn-ions changes with the Mn-Mn distance, we plot the absolute value of energy difference between AFM and FM alignments of the magnetic moments of Mn-atoms as a function of theirs separation (see Fig. \ref{fig:recon}). We assume collinear magnetic configurations of Mn-ions in which magnetic moments are either parallel or antiparallel. We do not include spin-orbit interaction (SOI), because it has been reported recently, that the SOI have a small influence on exchange energy and on the exchange coupling constant J (few meV) \cite{PhysRevB.85.155306}.

Since the exchange interaction is a crucial quantity in the field of DMS, we start our discussion comparing our results with the other theoretical predictions reported in literature \cite{PhysRevB.85.155306,PhysRevB.81.054401,PhysRevLett.93.177201,PhysRevB.70.235209} (see Fig. \ref{fig:recon}), for both  bulk and surface calculations obtained within different methods and different concentration of Mn-ions. We would like to stress that the two older surface calculations \cite{PhysRevB.81.054401,PhysRevB.85.155306} dealt with non-polar [110] GaAs surface that exhibits pronounced differences in comparison to the surfaces studies here.

The general trends of the exchange energy of Mn-pair at surfaces and in the bulk are similar, generally exhibiting a decay of the magnetic interaction with increasing distance between the magnetic ions.

As it is clearly seen in Fig. \ref{fig:recon}, our results of the exchange energy are in the same range of energy as the rest DFT calculations depicted in this figure (for both bulk and surface). The largest discrepancy is for the Tight Binding (T-B) calculations (see light pink curve with squared data points), for which the exchange energy values are approximately two times larger than energies obtained in DFT calculations. It is worth to mention, that our results for exchange energy for the given distance between the Mn-atoms constituting the pair are scattered within some energy range (indicated in Fig. \ref{fig:recon} by bars) that is of order of 100 meV for the nearest and second nearest distances between Mn-atoms. This is caused by the fact that at reconstructed surfaces the two Mn-ions in given distance between them can be placed in series of nonequivalent ways. We would like to emphasize, that the small energy difference of 13 meV (seen in Fig. \ref{fig:recon}) in the case of exchange energy obtained within the T-B calculations for the nearest distance between Mn-atoms results from the SOI taken into account in the one of the T-B computations. This clearly demonstrates that the SOI effect on the exchange energy is an order of magnitude weaker than the ‘environmental’ effect observed in our calculations and the neglect of SOI effects is justified for our purposes. In the following subsection, we present the detailed studies of the exchange coupling constants $J$ which exhibit some kind of anisotropy, i.e., the dependence on the crystallographic direction along which the Mn-pair is placed.

\subsubsection{Effective exchange coupling}

We have analyzed the effective exchange constant $J$ for the all non-equivalent NN distance between the Mn-ions at reconstructed (001) GaAs surfaces: $(2\times1)$, $\beta(2\times4)$, and $\beta2(2\times4)$ (see table \ref{tab:exchange1}), defined according to the formula: $E_{AFM}-E_{FM}=2JS^2$.
Here, we assume the magnetic spins of Mn-atoms to be $S=\frac{5}{2}$.
\begin{table}
\small
\caption{The exchange coupling $J$ given in meV for the nearest neighbor Mn-pairs for various considered reconstructed surfaces.
 The numbers in brackets delimit the obtained range of the exchange coupling.
The results are obtained employing standard L(S)DA and L(S)DA+U (for $U = 4.5 \, \mathrm{eV}$) approaches.} 
\def\arraystretch{1.4}
\label{tab:exchange1}
 \begin{tabular*}{\columnwidth}{@{\extracolsep{\fill}}lccc}
    \hline\vspace{1.5pt}
$J$ [meV] & $(2\times1)$&  $\beta(2\times4)$& $\beta2(2\times4)$  \\
\hline
&\multicolumn{3}{c}{Mn-Mn pair along $[110]$} \\
L(S)DA &  +12 & $[+12, +24]$&  $+16$\\
   L(S)DA+U  &+7 &$[+8, +19 ]$& $+13$\\
\hline
&\multicolumn{3}{c}{Mn-Mn pair along $[1\bar{1}0]$} \\
 L(S)DA  &  $[-16, -13]$& $[-10, +4]$& $[-3, +8]$\\
   L(S)DA+U& $[-18, -16]$& $[-7, +1]$& $[-7, +6]$\\\hline
  \end{tabular*}
\end{table}

Our results clearly demonstrate that the Mn-NN-pair along [110] crystallographic direction exhibits ferromagnetic alignment of spins, i.e., positive values of the exchange coupling $J$ (see table \ref{tab:exchange1}), with magnetization vectors on each Mn-atom being oriented along the [001] direction and of magnitude equal to 4.7 $\mu_B$, independently of the type of reconstructed surface. For the Mn-pair along [1$\bar{1}$0] crystallographic direction, the anti-ferromagnetic alignment of the spins is most likely to appear (mostly negative values of the exchange coupling $J$, see table \ref{tab:exchange1}). In other words, the mechanism of the magnetic ordering has the anisotropic character, namely it depends on the Mn-pair orientation. This rises a question whether it would be possible (e.g., by the STM method) to incorporate on purpose the magnetic atoms along a given direction at the surface. Then one could obtain on purpose magnetic or non-magnetic material. One can see that the value of the exchange constant for a given orientation of Mn-pair depends on the type of surface reconstruction, indicating that the exchange constant depends sensitively on the lattice arrangement of the atoms.

Now we compare our result with the exchange coupling for the pair reported so far in the literature. Those reports show that the effective $J$ for the pair is highly sensitive to doping levels and increases with decreasing Mn concentration \cite{PhysRevLett.93.177201,PhysRevB.70.235209,PhysRevB.63.233205,PhysRevB.69.115208}. The order of magnitude of the exchange coupling calculated in our studies is the same as previously reported for the surface and bulk results. However, the bulk calculations always predict the ferromagnetic ordering of the Mn-spins.

To our knowledge, previous surface calculations considered only the unreconstructed (110) non-polar GaAs surface. The authors of Ref. \cite{PhysRevB.85.155306} obtained, within the GGA approximation of DFT, a $J$ value equal to 17.9 meV \footnote{See Table VII in Ref. \cite{PhysRevB.85.155306}. For $U = 4 \, \mathrm{eV}$, the value of the difference between the $E_{AFM}$ and $E_{FM}$ is equal to 223.7 meV. In order to obtain the exchange constant $J$, we normalized the values quoted in Ref. \cite{PhysRevB.85.155306} dividing them by the factor $2 \cdot (5/2)^2=12.5$ (see the formula for the exchange coupling).} for the nearest neighbor configuration. Strandberg \textsl{et al.} \cite{PhysRevB.81.054401} obtained positive values of $J$ for two different directions of Mn-NN-pair at the (110) surface, 54 meV and 53 meV, for the very low Mn concentration of $x=0.0006$. They used the kinetic tight binding model. Due to the restrictions of the model they did not take into account relaxations of the atoms at the surface. This can significantly influence the results. Therefore, we believe that our studies provide reliable quantitative theoretical predictions and shed light on physical mechanisms leading to magnetic structure of Mn-pairs on the (001) GaAs surfaces. Further, we corroborate that the ordering mechanism of Mn-spins is governed by the local environment of Mn-atoms.

\subsubsection{Electronic structure and spin magnetic moments}

\begin{figure}
\includegraphics[width=0.95\columnwidth]{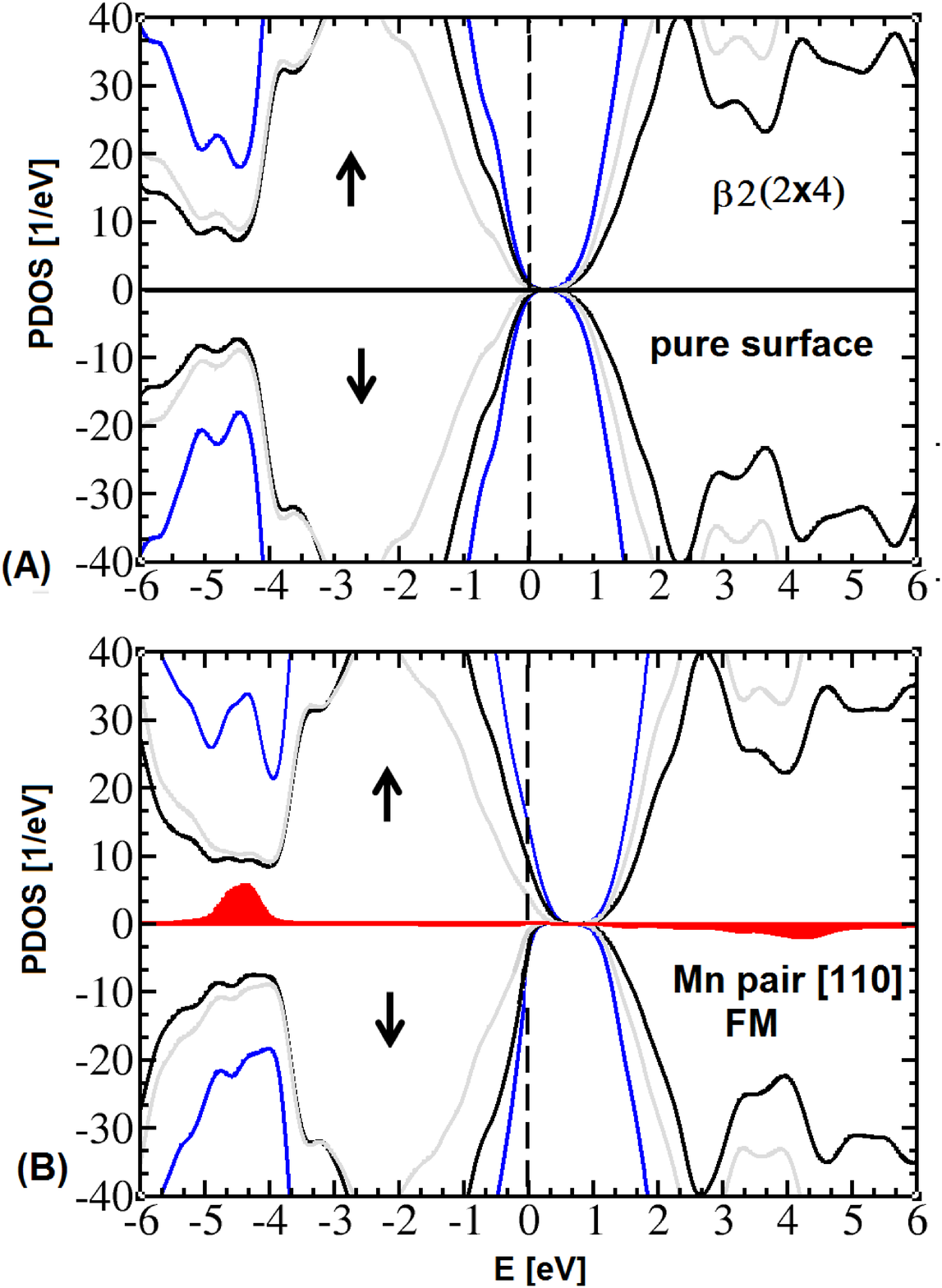}
\caption{\label{fig:pure}  Projected DOS  calculated within L(S)DA+U method for: (A) pure $\beta2(2\times4)$, and (B) the Mn-NN-pair substituted onto cationic sublattice along [110]  crystallographic direction at the $\beta2(2\times4)$ surface. The pure $\beta2(2\times4)$ surface is semiconducting, whereas substituted Mn-ions introduce extra states above the Fermi level within the valence band. The vertical dashed line denotes the position of the Fermi level. The solid blue, black, and green lines correspond respectively to total density of all states in a system, total density of As-states, and total density of Ga-states. The red area is the contribution from Mn $d$ states. The positive values of PDOS represent spin-up channel, negative spin-down one.}
\end{figure}

Let us discuss now the changes of electronic structure caused by incorporation of Mn-NN-pair at the reconstructed (001) GaAs.

\begin{figure}
\centering
\includegraphics[width=0.9\columnwidth]{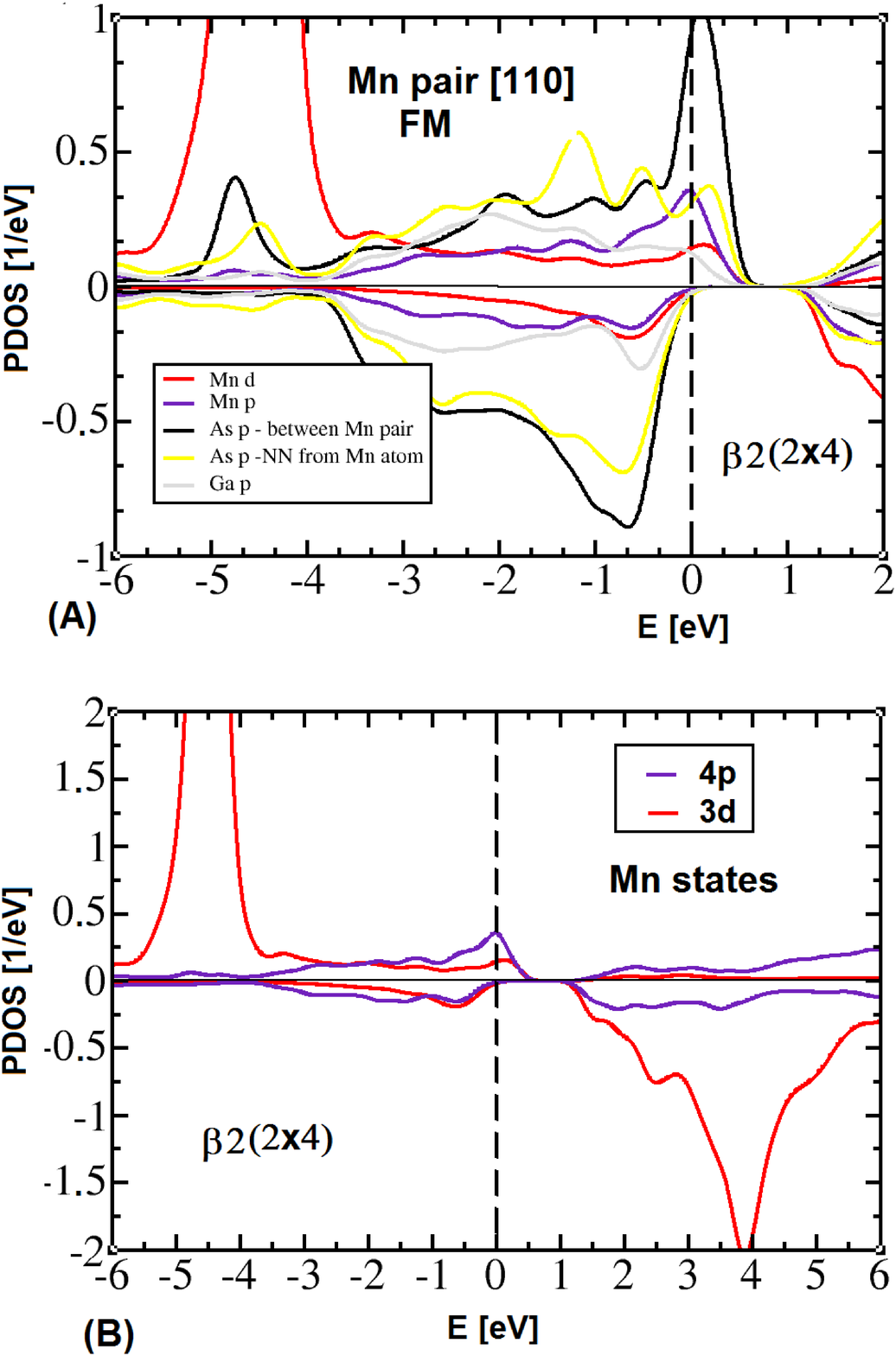}
\caption{\label{fig:Mnpdos} PDOS as obtained from the L(S)DA+U calculations for the Mn-NN-pair at the reconstructed $\beta2(2\times4)$ surface. (A) Contribution of the surrounding atoms of the Mn-NN-pair along the [110] crystallographic direction, and (B) the comparison between the Mn $p$ and $d$ states.}
\end{figure}

The pure $(2\times1)$ surface is metallic and the surface states are placed in the bandgap of the bulk, whereas the pure $\beta$-surfaces are semiconducting (see Fig. \ref{fig:pure}(A)) and do not introduce the surface states into the bulk's gap. Here, we focus on the electronic structure of the $\beta2(2\times4)$ reconstructed surface with the incorporated Mn-NN-pair. When the two Ga-ions are substituted by the Mn-ions, the extra electronic states appear just above the Fermi level (see Fig. \ref{fig:pure}(B)).
 These empty states which are above the Fermi level and belong to the valence band can be identified with the hole states. As hole states, we consider unoccupied states with energies between Fermi energy (lying in the valence band)\footnote{We have not considered the holes in the system with Mn-NN-pair on the $(2\times1)$ reconstructed surface. The reason of that is the appearance of the unoccupied surface states in the bandgap for the pure $(2\times1)$ reconstruction, and theirs mixing with the unoccupied states coming from the substitution of the Mn-atoms at cationic sublattice.} and top of the valence band, as it was previously considered for a bulk system \cite{PhysRevB.69.195203,Milowska20147}. The hole states have mostly $p$ character (see Fig. \ref{fig:Mnpdos}(A)).

 The greatest contribution to the holes comes from the arsenic
 atoms that are the nearest neighbors of the Mn-NN-pair and reside along the [110] crystallographic direction (see Fig. \ref{fig:Mnpdos}(A)). Moreover, the $3d$ states are mainly localized around 4.5 eV below the Fermi level (see Fig. \ref{fig:Mnpdos}(B)). Therefore, the Mn $p$ states hybridize with the surrounding stronger than the $3d$ states. There is only small admixture of the Mn $d$ states with the hole states.

Now let us discuss the spatial distribution of the hole, in order to visualize its character.
\begin{figure}
\centering
\includegraphics[width=0.85\columnwidth]{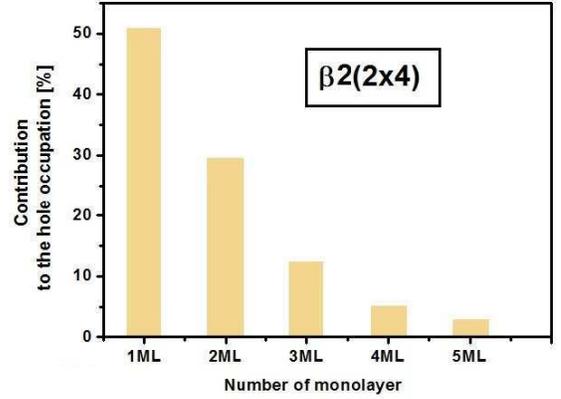}
\caption{\label{fig:Holesb22x4} Contribution [in \%] to the hole occupation coming from various layers of the slab representing $\beta2(2\times4)$ reconstructed (001) GaAs surface with incorporated Mn-NN-pair along the [110] direction. 1ML indicates the top layer of As-atoms, 2ML indicates the Ga-atoms layer just below the top As-layer. The Mn-NN-pair is placed in this layer. 3ML \& 5 ML one As-layers, 4ML is Ga-layer. The higher the number the deeper position of the layer relative to the top one (1ML).}
\end{figure}

The hole occupation $N_h$ is defined here as the integral over energy of the density of states from the Fermi energy to the top of the valence band. Note that the same procedure was previously adopted in Refs. \cite{PhysRevB.69.195203} and \cite{Milowska20147} for the bulk system. One can also consider integrated density of states $N_{h,layer}$ coming from the projected DOS for various slab layers. The analysis of the contribution of the various layers to the total hole density ($N_h$) defined above is presented in Fig. \ref{fig:Holesb22x4} in the case of the discussed here $\beta2(2\times4)$ reconstructed surface.

It is clearly seen that the contribution to the total hole occupancy decreases quickly with the depths of the layer, and the highest two layers (indicated as the 1ML \& 2ML in Fig. \ref{fig:Holesb22x4}) contribute 80\% of the hole density. This result sheds light on the degree of the hole delocalization. We would like to emphasize that in our calculations no impurity band related to surface acceptor has been observed. This clearly corresponds to the Zener type (or RKKY one) \cite{Science.2000} of magnetic coupling between Mn-ions, and the picture of surface, ligand-ion, $p-d$ interaction is the adequate one.
\begin{figure}
\centering
\includegraphics[width=1.\columnwidth]{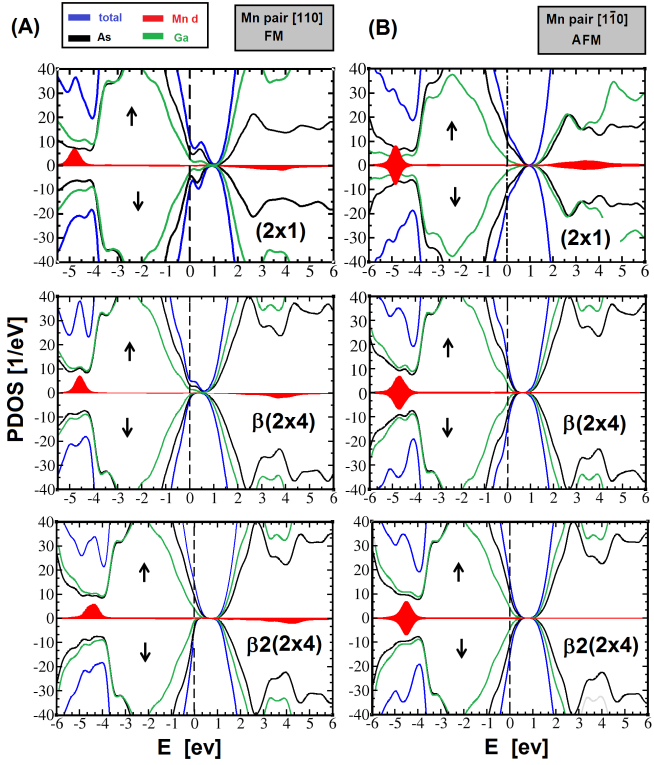}
\caption{\label{fig:pdos} PDOS calculated within L(S)DA+U
method for the three reconstructed surfaces of GaAs with Mn-NN-pair along two different orientations: [110] (left side of the picture) and [1$\bar{1}$0] (right side of the picture). Spin alignments for Mn-NN-pairs are FM and AFM, for direction [110] and [1$\bar{1}$0], respectively. The vertical dashed line denotes the position of the Fermi level. The solid blue, black, and green lines correspond respectively to total density of all states in a system, total density of As-states, and total density of Ga-states. The red area indicates the contribution from Mn $d$ states.}
\end{figure}

The electronic structure of the $\beta(2\times4)$ reconstructed (001) GaAs surface with the Mn-NN-pair with FM ordering of spins placed along [110] direction is typical for other considered reconstructed (001) GaAs surfaces with incorporated Mn-NN-pairs and exhibits all essential features. It is illustrated in Fig. \ref{fig:pdos}(A). For comparison, in Fig. \ref{fig:pdos}(B), we plot also density of states for the Mn-NN-pair with AFM spin ordering incorporated along the [1$\bar{1}$0] direction into three types of the reconstructed GaAs (001) surfaces discussed in the present paper. With the discussion presented above, this figure is self-explanatory. 

\begin{figure}
\centering
\includegraphics[width=0.95\columnwidth]{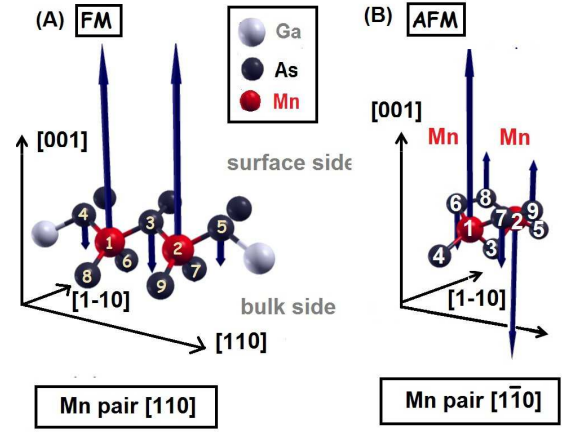}
\caption{\label{fig:MomentMn110} Distribution of magnetic moments on As-atoms surrounding the Mn-NN-pair placed along [110] (A) and [1$\bar{1}$0] (B) crystallographic directions at the $(2\times1)$ reconstructed (001) GaAs surface. The blue arrows indicate the magnetic moments of Mn-atoms with FM (A) and AFM (B) alignments. The lengths of the magnetic vectors on As-sites are exaggerated for clarity of the presentation.}
\end{figure}

Now, let us focus on distribution of magnetic moments in the surrounding of the Mn-NN-pair by using Mulliken analysis \cite{Mulliken.1955}. Our results clearly show that the local magnetic moments of Mn-ions polarize their surrounding always in such a way, that the nearest neighbor As-atoms acquire the magnetic moments anti-ferromagnetically aligned to the Mn-spins, which is schematically illustrated in Fig. \ref{fig:MomentMn110} for the case of $(2\times1)$ reconstructed surface. For Mn-NN-pair along [110] direction (i.e., with FM ordering of spins), the values of magnetic moments on As-atoms surrounding Mn ones are listed for three considered surface reconstructions in Table \ref{comp}.

\begin{table}
\small
\caption{The spin magnetic polarization (in $\mu_B$ per atom) of the As-atoms surrounding Mn-ions for the FM arrangement of Mn-NN-pair at $[110]$ direction for various reconstructed surfaces, as calculated by using Mulliken analysis \cite{Mulliken.1955}. The As-atoms are numbered according to the Fig. \ref{fig:MomentMn110}.}
\label{comp}
\begin{centering}
\def\arraystretch{1.4}
\begin{tabular*}{\columnwidth}{@{\extracolsep{\fill}}lccc}
\hline
   &\multicolumn{3}{c}{ Surface reconstruction } \\
atom &  $(2\times1)$ &  $\beta(2\times4)$ & $\beta2(2\times4)$   \\
\hline
 As3 & $-0.74$ & $-0.75$ & $-0.79$    \\
 As4 & $-0.2$ & $-0.17$ & $-0.2$     \\
 As5 & $-0.2$ & $-0.17$ &  ---            \\
 As9 = As7& $-0.12$& $-0.10$& $-0.17$\\
 As8 = As6& $-0.12$& $-0.10$& $-0.09$\\
\hline
  \end{tabular*}
\end{centering}
\end{table}

One can see that the induced magnetic moments are comparable and do not depend on the type of reconstructed surfaces. They decrease rapidly with the increasing distance to the Mn-pair. Only the nearest neighbors of the Mn-pair have (i.e., these depicted in Fig. \ref{fig:MomentMn110}) noticeable induced spin polarization. For the second neighbors, the induced magnetic moments are smaller than 0.01 $\mu_B$, and for the third neighbors, the magnetic moments are nearly equal to zero. The same picture emerges from both L(S)DA and L(S)DA+U calculations. In the bulk system, the magnetic moment on the first, second, and third Mn neighbor extends to 0.11 $\mu_B$, 0.015 $\mu_B$, and 0.015 $\mu_B$, respectively (according to the L(S)DA+U \cite{PhysRevB.70.235209}). This comparison between the bulk and the surface shows that the induced magnetic moments are more localized at the surface than in the bulk. In addition, the induced magnetic moments on the atoms are always larger at the surface than at the monolayer closer to the bulk site. In the case of AFM alignment of Mn-spins, the As-atom positioned between the Mn-NN-pair is not spin polarized. Other As neighbors of Mn-atoms have similar absolute values of spin polarization as in the FM case.

\section{Conclusions}
\label{conclude}
We have studied the stability, morphology, and electronic structure of Mn-pairs substituted onto Ga sublattice at different As-terminated reconstructed (001) GaAs surfaces. We have demonstrated that the energy of the system depends on the position, orientation, and the distance between the Mn-atoms. The Mn-pairs with Mn-atoms being the nearest neighbors (i.e., constituting Mn-pair) are energetically the most favorable, independently on the surface reconstruction pattern. This causes that there are only two crystallographic directions, namely [110] and [1$\bar{1}$0] relevant for incorporation of Mn-NN-pair at the surface. For $(2\times1)$ and $\beta(2\times4)$ reconstruction patterns, the Mn-NN-pair placed along [110] direction is more stable than Mn-NN-pair placed along [1$\bar{1}$0] direction. On the contrary, for $(1\times1)$ and $\beta2(2\times4)$ reconstructions, the energetically stable configuration of the system requires Mn-NN-pair along the [1$\bar{1}$0] direction. Our results point out to a possibility of affecting the magnetic anisotropy of the (Ga,Mn)As layer \cite{PhysRevLett.108.237203} by forcing an alternative reconstruction of the semiconductor surface, during the growth, by choosing appropriate growth conditions - namely the vapor pressure, temperature or surface composition \cite{Shimizu.1999,Schott.2001}. Moreover, we have demonstrated that the mechanism of the magnetic ordering depends on the Mn-NN-pair orientation. Generally, the Mn-NN-pair along [110] crystallographic direction exhibits ferromagnetic alignment of the spins, whereas the Mn-NN-pair along [1$\bar{1}$0] crystallographic direction prefers to be anti-ferromagnetically aligned. 
In addition, we have shown that the Mn $4p$ states have greater contribution to the hole states than $3d$ states, which is consistent with recently published theoretical results for bulk (Ga,Mn)As \cite{PhysRevB.69.195203,Milowska20147}.

We have demonstrated that the holes are mostly localized in the closest proximity of the Mn-NN-pair. Nevertheless, there are small contributions which come from the distant atoms indicating extended character of the holes. These observations strongly suggest that the Zener $p-d$ interaction \cite{Science.2000} can also cause ferromagnetic coupling of the Mn-NN-pairs at the surface, provided the magnetic moments of the Mn-atoms forming the pair do not cancel each other
(FM configurations).

Furthermore, our results clearly show that the local magnetic moments of Mn-ions polarize their surrounding always in such a way, that the Mn nearest neighbor As-atoms acquire the magnetic moments which are anti-ferromagnetically aligned to the Mn-spins.






\begin{acknowledgments}
The authors are grateful to T. Dietl for illuminating discussions. This work was supported by the European Research Council through the FunDMS Advanced Grant within the ``Ideas'' Seventh Framework Programme of the EC and InTechFun (POIG.01.03.01-00-159/08) of EC. We made use of computing facilities of PL-Grid Polish Infrastructure for Supporting Computational Science in the European Research Space, and acknowledge the access to the computing facilities of the Interdisciplinary Centre of Modelling, University of Warsaw. The support of the National Research Council (NCN) through the grant HARMONIA DEC-2013/10/M/ST3/00793 is gratefully acknowledged.
\end{acknowledgments}
\bibliographystyle{apsrev4-1}
\bibliography{bibliografia}

\end{document}